\documentclass{ws-ijmpc}

\begin{document}

\markboth{Jian-Guo Liu et. al.} {Effects of the users taste on
personalized recommendation}

\catchline{}{}{}{}{}

\title{EFFECTS OF USER TASTES ON PERSONALIZED RECOMMENDATION}

\author{JIAN-GUO LIU, TAO ZHOU, BING-HONG WANG, YI-CHENG ZHANG}

\address{Research Center of Complex Systems Science, University
of Shanghai for Science and Technology, Shanghai 200093, P.R. China\\
Department of Modern Physics, University of Science and Technology
of China, 230026, P.R. China\\Department of Physics, University of
Fribourg, Fribourg CH-1700, Switzerland\\liujg004@ustc.edu.cn,
zhutou@ustc.edu}

\author{QIANG GUO}
\address{School of Science, University of Shanghai for Science
and Technology, Shanghai 200093, P.R. China}

\maketitle

\begin{history}
\received{Day Month Year}
\revised{Day Month Year}
\end{history}

\begin{abstract}
In this paper, based on a weighted projection of the user-object
bipartite network, we study the effects of user tastes on the
mass-diffusion-based personalized recommendation algorithm, where a
user's tastes or interests are defined by the average degree of the
objects he has collected. We argue that the initial recommendation
power located on the objects should be determined by both of their
degree and the users' tastes. By introducing a tunable parameter,
the user taste effects on the configuration of initial
recommendation power distribution are investigated. The numerical
results indicate that the presented algorithm could improve the
accuracy, measured by the average ranking score, more importantly,
we find that when the data is sparse, the algorithm should give more
recommendation power to the objects whose degrees are close to the
users' tastes, while when the data becomes dense, it should assign
more power on the objects whose degrees are significantly different
from user's tastes.

\end{abstract}

\ccode{PACS Nos.: 89.75.Hc, 87.23.Ge, 05.70.Ln}

\begin{figure}[b]
\center\scalebox{0.8}[0.8]{\includegraphics{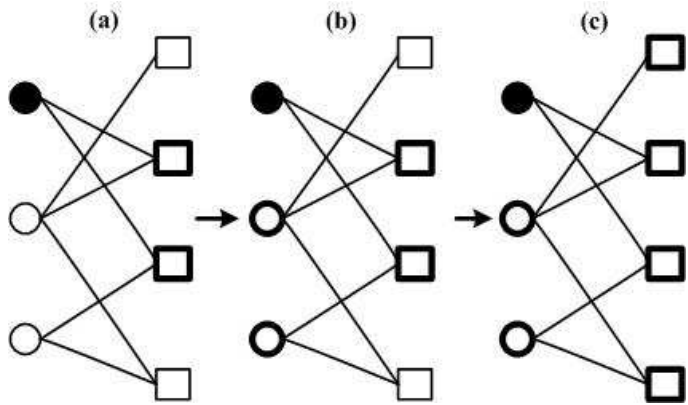}}
\caption{Illustration of the network-based algorithm. The
network-based algorithm could be applied in the following way. (a)
The objects collected by the target user are activated; (b) The heat
is diffused from the activated objects to the users who have
collected them; (c) Then it's diffused back from the users to the
objects.}\label{Fig1.1}
\end{figure}

\section{Introduction}
With the rapid growth of the Internet and the World-Wide-Web, a huge
amount of data and resource confront people with an information
overload \cite{Broder2000}. There are thousands of movies, millions
of books, and billions of web pages on the web sites, and the amount
of information is increasing more quickly than our personal
processing abilities. This brings about massive amount of accessible
information, which may result in a dilemma problem. It's hard for us
to effectively filter out the pieces of information that are most
appropriate for us. A landmark for information filtering is the use
of search engine \cite{3,4}, by which user could find the relevant
web pages by putting certain keywords. However, the search engine
only returns the same results regardless of the users' tastes and
interests.

Thus far, the most promising way to efficiently filter out the
information overload is to provide {\it personalized
recommendations}, which attempts to find out objects likely to be
interesting to the target users by extracting the hidden information
from the users' historical selections or collections. Motivated by
its significance for economy and society, the design of efficient
recommendation algorithms has become a common focus for computer
science, mathematics, marketing practices, management science and
physics. Various kinds of algorithms have been proposed, such
collaborative filtering (CF) approaches
\cite{Herlocker2004,Konstan1997,Liu2009,LiuRR2009,Duo2009},
content-based analyses \cite{Balab97,Pazzani99}, network-based
algorithm \cite{Zhang2007,Zhang2007b,Zhou2007a,Zhou2007b}, hybrid
algorithms \cite{Good1999,Pazzani1997}, and so on. For a review of
current progress, see Refs. \cite{Adomavicius2005,Liu2009b} and the
references therein.

Very recently, some physical dynamics, including mass diffusion (MD)
\cite{Zhou2007a,Zhou2007b} and heat conduction (HC)
\cite{Zhang2007}, have found their applications in personalized
recommendations. These algorithms have been demonstrated to be of
both high accuracy and low computational complexity
\cite{Zhang2007,Zhang2007b,Zhou2007a,Zhou2007b}. Since MD and HC
algorithms could be implemented based on the user-object bipartite
network, it's also called {\it network-based algorithm}. The
network-based algorithm supposes that the objects one user has
collected have the power to recommend new objects to the target
user, which is coincidence with the definition reachability
\cite{PRE76}. In this paper, we introduce an improved MD algorithm
with user-taste-dependent initial configuration. Compared with the
uniform initial configuration, the prediction accuracy can be
enhanced by using the user-taste-dependent configuration. More
significantly, besides the prediction accuracy, we find that the
data sparsity is an important factor affecting the algorithm
performance. When the sparsity of the user-object bipartite network
is small, in other words, there are few edges between the users and
objects, the algorithm should pay more attention on the users'
habits and tastes, while when the number of edges in the bipartite
network is large, the algorithm should give more recommendation
power on the objects whose degrees significantly different with
users' habits. Numerical simulations show that the improved
algorithm has higher accuracy and can provide more diverse and less
popular recommendations.

\section{Mass-diffusion-based personal recommendation}
In a recommender system, each user has voted or collected some
objects, the system could be described by a bipartite network, in
which there are two kind of nodes users and objects, the users'
historical collection or selection behaviors could be well
demonstrated by the edges connecting the users and objects.
Formally, denote the object set as $O = \{o_1,o_2, \cdots, o_m\}$
and the user set as $U$ = $\{u_1, u_2,$ $\cdots,$  $u_n\}$, the
system can be fully described by a bipartite network with $m+n$
nodes, where there is an edge between a user and object if and only
if this object is collected by the user. The bipartite network could
be described by an adjacent matrix ${\bf A}=\{a_{ij}\}\in R^{m,n}$,
where $a_{ij}=1$ if $o_i$ is collected by $u_j$, and $a_{ij}=0$
otherwise. In MD algorithm, an object-object similarity network
${\bf W}=\{w_{\alpha\beta}\}_{m,m}$ is constructed firstly, where
each node represents an object and two objects are connected if they
have been collected simultaneously by at least one user. Then, to a
target user, an amount of recommendation power is set on each object
he has collected, and the proportion of the resource
$w_{\alpha\beta}$ would like to distribute from $o_\beta$ to
$o_\alpha$. In MD, a reasonable assumption is that the objects that
users have collected are what they like, and the objects a target
user has collected would be regarded as the initial mass source,
then the activated objects redistribute the mass to the users who
have collected them before, with users receiving a level of mass
equal to the mean amount possessed by their neighboring objects, and
objects then receiving back the mean of their neighboring users¡¯
mass levels. Due to the sparsity of real data sets, these
``physical" descriptions of the algorithm turn out to be more
computationally efficient in practice than constructing and using
the object similarity matrix ${\bf W}$, and MD algorithm could be
implemented in three steps on the user-object bipartite network,
which is shown in Fig.\ref{Fig1.1}(a-c).

Lind {\it et. al.} presented a cycle measurement to investigate the
clustering property in bipartite network \cite{Han1,Han2}. According
to the algorithm description and the cycle definition, the object
similarity of the mass-diffusion-based algorithm can be expressed as
\cite{Zhou2007a},
\begin{equation}
w_{\alpha\beta}=\frac{1}{k(o_{\beta})}\sum_{l=1}^n\frac{a_{\alpha
l}a_{\beta l}}{k(u_l)},
\end{equation}
where $k(o_{\beta})=\sum_{i=1}^na_{\beta i}$ and
$k(u_l)=\sum_{i=1}^m a_{il}$ denote the degrees of object
$o_{\beta}$ and user $u_l$, respectively. For a target user, in the
simplest case, the initial resource vector ${\bf
f}=\{f_1,f_2,\cdots,f_m\}^T$ can be set as
\begin{equation}
f_j=a_{ji}.
\end{equation}
In other words, only the objects user $u_i$ has collected are set
unit resource. After the mass-diffusion-process demonstrated in
Fig.1, the final resource vector is
\begin{equation}
{\bf \widehat{f}}=W{\bf f}.
\end{equation}
Sorting the vector $\bf \widehat{f}$ in descending order according
to value of $\widehat{f}_j$, the objects obtained highest values are
recommended to the target user.

\begin{figure}[t]
\center\scalebox{0.4}[0.4]{\includegraphics{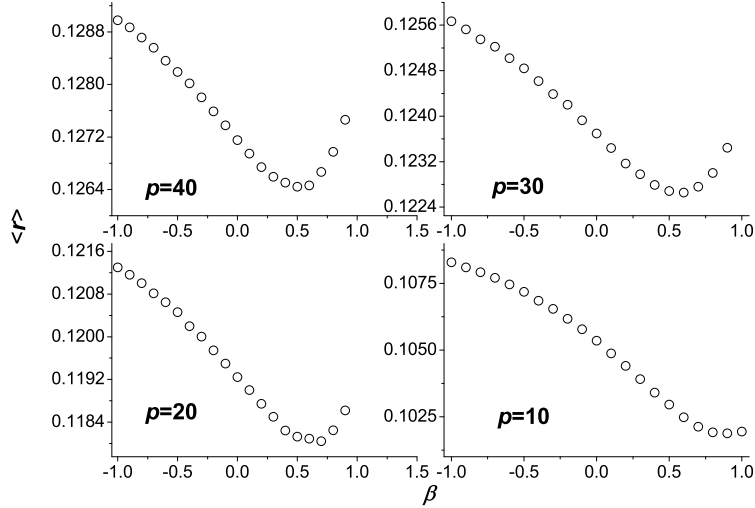}}
\caption{Average ranking score $\langle r\rangle$ vs. $\beta$ when
$p=10,20,30$ and 40. All the data points are averaged over ten
independent runs with different data-set divisions.}\label{Fig1}
\end{figure}

\section{Improved algorithm by considering the user taste effects}
In the standard MD algorithm, for any user $u_i$, all of the
collected objects are assigned the same recommendation power.
Although it already has a good algorithmic accuracy, this uniform
configuration may be oversimplified, and didn't consider the effects
of user tastes. In this paper, the user taste is defined by the
average object degree he has collected. The objects whose degrees
are close to the user taste should be assigned more recommendation
power. We also notice that most of the user tastes are less than
100, while the degrees of the popular objects are close to 300. If
the recommendation power is assigned according to the distance
between the object degree and the user taste, it will give more
power on the popular objects and weaken the unpopular object
effects. In order to balance the objects whose degrees are larger or
less than the user tastes, we present a more complicated
distribution of initial resource according to the following way.
\begin{equation}
f^i_{\alpha} = a_{\alpha i}I_{\alpha i},
\end{equation}
where $I_{\alpha i}$ is defined as follows
\begin{equation}
I_{\alpha i}=\left\{
\begin{array}{cl}
(k(o_{\alpha})/\overline{k}(u_i))^\beta  &
k(o_{\alpha})\geq \overline{k}(u_i) \\[5pt]
(\overline{k}(u_i)/k(o_{\alpha}))^\beta  &
k(o_{\alpha})<\overline{k}(u_i)
\end{array}
\right.
\end{equation}
where $\overline{k}(u_i)$ denote the average degree of user $u_i$'s
collected objects, and $\beta$ is a tunable parameter. Compared with
the uniform case, $\beta=0$, a positive $\beta$ strengthens the
influence of the objects whose degrees are larger or less than
$\overline{k}(u_i)$, while a negative $\beta$ strengthen the
influence of the objects whose degrees are close to
$\overline{k}(u_i)$.

\begin{figure}[t]
\center\scalebox{0.4}[0.4]{\includegraphics{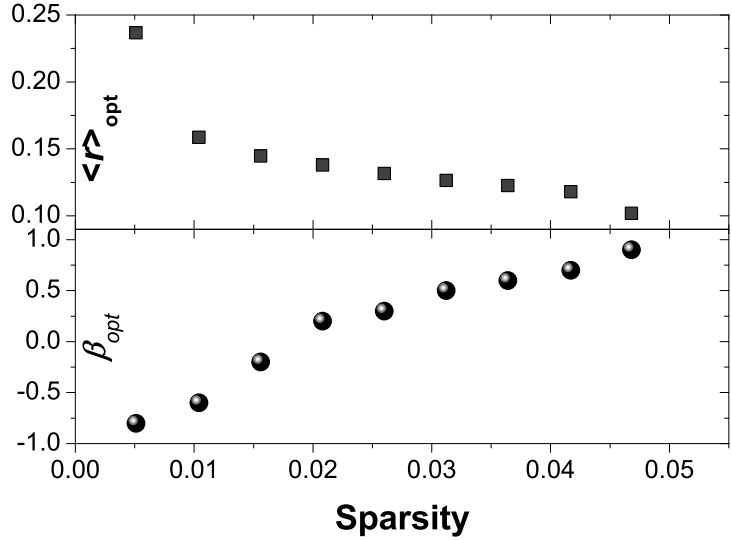}} \caption{The
optimal $\beta_{\rm opt}$ and the corresponding average ranking
score $\langle r\rangle_{\rm opt}$ vs. the sparsity of the training
set. All the data points are averaged over ten independent runs with
different data-set divisions.}\label{Fig2}
\end{figure}

\section{Numerical results}
A benchmark dataset, namely MovieLens, is used to test the improved
algorithm. The MovieLens data is a randomly-selected subset of the
huge data, which consists of 1682 movies (objects) and 943 users.
The users vote movies by discrete ratings from one to five. We
applied a coarse-graining method: A movie is set to be collected by
a user only if the giving rating is larger than 2. The original data
contains $10^5$ ratings, 85.25\% of which are $\geq 3$, that is, the
user-object (user-movie) bipartite network after the coarse gaining
contains 85250 edges. We randomly divide this data set into two
parts: one is the training set, treated as known information, and
the other is the probe, whose information is not allowed to be used
for prediction. We use a parameter $p$ to control the data density,
that is, $p$\% of the ratings are put into the probe set, and the
remains compose the training set.

A good recommender method should rank preferable objects to math the
users' tastes. Therefore, the collected objects in the probe set
should be set at the top level of the recommendation lists. The
average ranking score is adopted to measure the accuracy. It could
be defined as follows. For a target user $u_i$, if the entry
$u_i$-$o_j$ is in the probe set, we measure the position of $o_j$ in
the ordered list. For example, if there are $L_i=10$ uncollected
objects for $u_i$, and $o_j$ is the 2nd one from the top, we say the
position of $o_j$ is $2/10$, denoted by $r_{ij}=0.2$. A good
algorithm is expected to give high recommendations to them, thus
leading to small $r_{ij}$. Therefore, the mean value of the position
$\langle r\rangle$ can be used to evaluate the algorithmic accuracy:
the smaller the average ranking score, the higher the algorithmic
accuracy, and vice verse. The average degree of all recommended
objects, $\langle k\rangle$, and the mean value of Hamming distance,
$S$, are taken into account to evaluate the popularity and
diversity. The smaller average degree, corresponding to the
unpopular objects, are preferred since those small-degree objects
are hard to be found by users themselves. The diversity can be
quantified by the average Hamming distance, $S=\langle
H_{ij}\rangle$, where $H_{ij}=1-Q_{ij}/L$, $L$ is the length of
recommendation list and $Q_{ij}$ is the overlapped number of objects
in $u_i$ and $u_j$'s recommendation lists. The largest $S=1$
indicates the recommendations to all of the users are totally
different, while the smallest $S=0$ means all of recommendations are
exactly same.

\begin{figure}[t]
\center\scalebox{0.4}[0.4]{\includegraphics{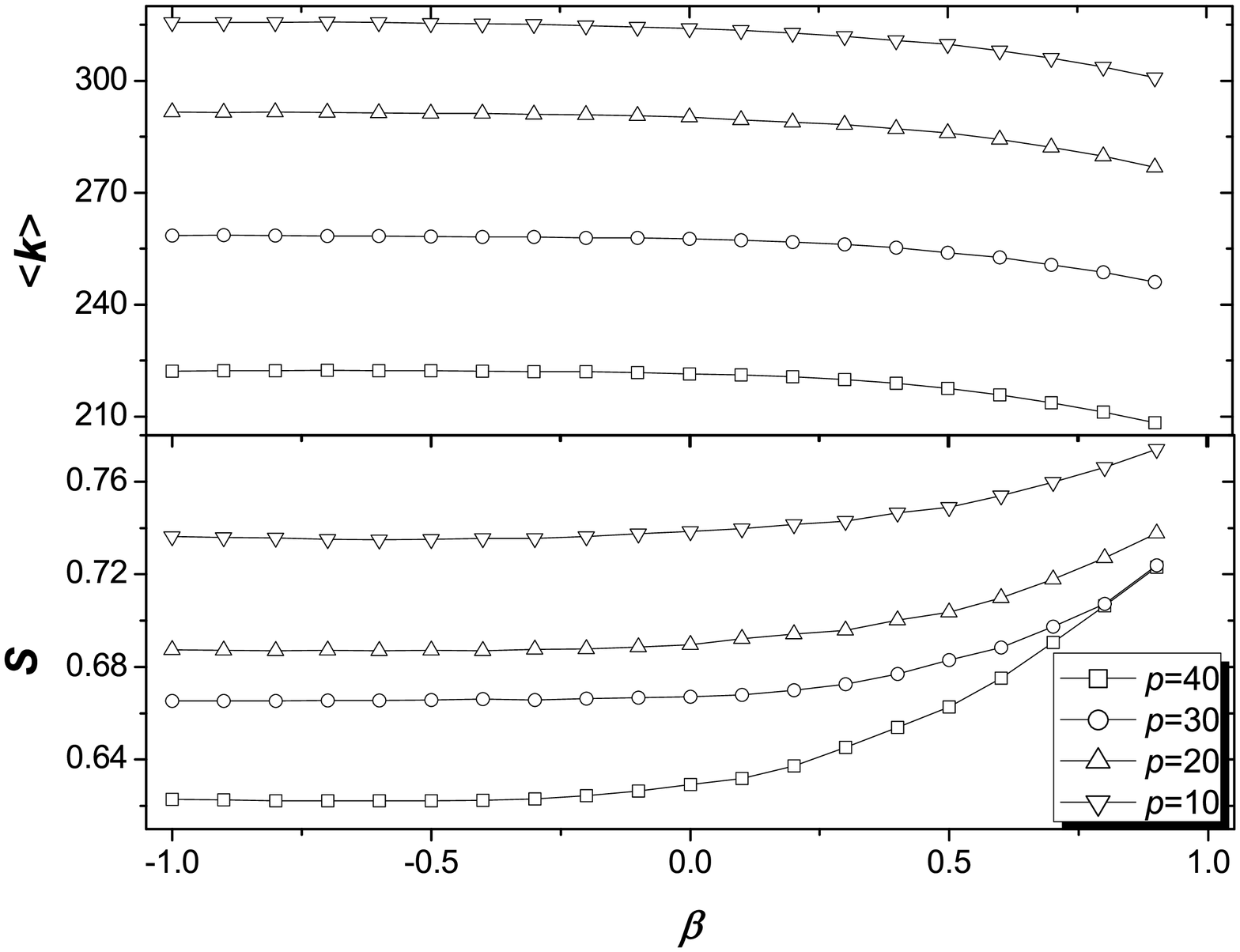}} \caption{When
the recommendation list $L=10$, $\langle k\rangle$ and $S$ vs.
$\beta$ of $p=10,20,30$ and 40. All the data points are averaged
over ten independent runs with different data-set
divisions.}\label{Fig3}
\end{figure}

Implementing the improved algorithm on the MovieLens data, the
accuracy, popularity and diversity are investigated. Figure
\ref{Fig1} reports the algorithmic accuracy as a function of $\beta$
to different $p$, from which one can find that the curves obtained
by the improved algorithm has clear minimums, which strongly support
the above discussion. Compared with the routine case ($\beta=0$),
the average ranking score can be reduced 5.6\% at the optimal case
when $p=10$. Numerical results on different percentage of probe sets
show that the optimal parameter $\beta_{\rm opt}$ decreases with the
increase of $p$. Figure \ref{Fig2} reports the relation between the
optimal $\beta_{\rm opt}$, the corresponding average ranking scores
$\langle r\rangle_{\rm opt}$ and the sparsity of the training sets.
One can see from Fig.\ref{Fig2} that the optimal $\langle
r\rangle_{\rm opt}$ is negatively correlated with the data sparsity,
where the sparsity is defined as $\frac{E}{m\times n}$, and $E$ is
the number of edges in the user-object bipartite network, more
interestingly, the optimal parameter $\beta_{\rm opt}$ is positively
correlated with the sparsity. The reason may lie in the fact that
when the users have not collected too much objects, their tastes are
easy to be distinguished, therefore, the objects whose degrees are
close to $\overline{k}(u_i)$ should be assigned more recommendation
power. As the number of users' collected objects increases, users'
tastes become diversity, therefore, it's hard to catch the users
interested and habits. Under these circumstances, the users are more
interesting to the objects different from his historical collects
which could bring him/her fresh information. Besides accuracy, the
popularity and diversity are also investigated. Figure \ref{Fig3}
reports the average degree and diversity of all recommended movies
as a function of $\beta$ to different $p$ when the recommendation
lists $L=10$, from which one can find that although the average
object degrees scarcely change, the diversity is increased at the
optimal $\beta_{\rm opt}$.

\section{Conclusion and Discussion}
In this paper, the effects of user tastes on MD recommendation
algorithm are investigated, where the user tastes are defined by the
average object degree he/her has collected. By introducing a free
parameter $\beta$, an improved algorithm by regulating the initial
configuration of resource is presented. Numerical results indicate
that when the data set is sparse, it's easy to distinguish the
users' tastes and the objects whose degrees are close to the users'
tastes should be assigned more recommendation power, while as the
data set becoming dense, the objects whose degree far from the
users' tastes should be emphasized. Besides the average ranking
score, the popularity and personalization of recommended objects are
also taken into account. The results show that the improved
algorithm outperforms the standard MD algorithm in both of accuracy
and personalization.

In the improved algorithm, we only give a kind of user taste
definition, however, there are several other ways to define the
users' tastes, such as time-dependent behavior, variance of the user
collected object degrees, and so on. We believe MD algorithm could
be further improved by catching the users' current tastes.

Instead of calculating all the elements in ${\bf W}$, one can
implement the current algorithm by directly diffusing the resource
of each user. Ignoring the degree-degree correlation in user-object
relations, the algorithmic complexity is $O(m\langle
k_u\rangle\langle k_o\rangle)$, where $\langle k_u\rangle$ and
$\langle k_o\rangle$ denote the average degrees of users and
objects. Theoretical physics provides us some beautiful and powerful
tools in dealing with this long-standing challenge in modern
information science: how to do a personal recommendation. The
presented algorithm also could be used to find the relevant
reviewers for the scientific papers or funding applications
\cite{Liu1,Liu2}, and the link prediction in social and biological
networks\cite{Zhou2009}. We believe the current work can enlighten
readers in this promising direction.

This work is partially supported by the National Basic Research
Program of China (No. 2006CB705500), the National Natural Science
Foundation of China (Nos. 60744003, 10635040, 10472116), the Swiss
National Science Foundation (Project 205120-113842), and Shanghai
Leading Discipline Project(Grant No. S30501).


\begin{thebibliography} {1}

\bibitem{Broder2000} G.-Q. Zhang, G.-Q. Zhang, Q.-F. Yang, S.-Q. Cheng and T.
Zhou, {\it New Journal of Physics} {\bf 10}, 12307 (2008).

\bibitem{3}S. Brin and L. Page, {\it Comput. Netw. ISDN Syst.} {\bf 30}, 107 (1998).

\bibitem{4}J.M. Kleinberg, {\it J. ACM} {\bf 46}, 604 (1999).

\bibitem{Herlocker2004}
J.L. Herlocker, J.A. Konstan, K. Terveen and J. Riedl, {\it ACM
Trans. Inform. Syst.} {\bf 22}, 5 (2004).

\bibitem{Konstan1997}
J.A. Konstan, B.N. Miller, D. Maltz, J.L. Herlocker, L.R. Gordon and
J. Riedl, {\it Commun. ACM} {\bf 40}, 77 (1997).

\bibitem{Liu2009}
J.-G. Liu, B.-H. Wang and Q. Guo, {\it Int. J. Mod. Phys. C} {\bf
20}, 285 (2009).

\bibitem{LiuRR2009}
R.-R. Liu, C.-X. Jia, T. Zhou, D. Sun and B.-H. Wang, {\it Physica
A} {\bf 388}, 462 (2009).

\bibitem{Duo2009}
D. Sun, T. Zhou, J.-G. Liu, R.-R. Liu, C.-X. Jia and B.-H. Wang,
{\it Phys. Rev. E} {\bf 80}, 017101 (2009).

\bibitem{Balab97}
M. Balabanovi\'c and Y. Shoham, {\it Commun. ACM} {\bf 40}, 66
(1997).

\bibitem{Pazzani99} M.J. Pazzani, {\it Artif. Intell. Rev.} {\bf 13}, 393 (1999).

\bibitem{Zhang2007} Y.-C. Zhang, M. Blattner and Y.-K. Yu, {\it Phys.
Rev. Lett.} {\bf 99}, 154301 (2007).

\bibitem{Zhang2007b}Y.-C. Zhang, M. Medo, J. Ren, T. Zhou, T. Li and F. Yang, {\it Europhys. Lett.} {\bf 80}, 68003 (2008).

\bibitem{Zhou2007a} T. Zhou, J. Ren, M. Medo and Y.-C. Zhang, {\it Phys. Rev. E} {\bf 76}, 046115 (2007).

\bibitem{Zhou2007b} T. Zhou, L.-L. Jiang, R.-Q. Su and Y.-C. Zhang, {\it Europhys. Lett.} {\bf 81}, 58004 (2008).

\bibitem{PRE76}
P.G. Lind, L.R. da Silva, J.S. Andrade and H.J. Herrmann, {\it Phys.
Rev. E} {\bf 76}, 036117 (2007).

\bibitem{Good1999}
N. Good, J.B. Schafer, J.A. Konstan, A.l. Borchers, B. Sarwar, J.
Herlocker and J. Riedl,
{\it Proc. Conf. Am. Assoc. Artificial Intelligence}, 439 (1999).



\bibitem{Pazzani1997}
M. Pazzani and D. Billsus, {\it Machine Learning} {\bf 27}, 313
(1997).

\bibitem{Adomavicius2005}G. Adomavicius and A. Tuzhilin, {\it IEEE Trans. Know. \& Data
Eng.}
{\bf 17}, 734 (2005).

\bibitem{Liu2009b}
J.-G. Liu, M.Z.Q. Chen, J. Chen, F. Deng, H.-T. Zhang, Z. Zhang and
T. Zhou. 
{\it Int. J. Info. and Syst. Sci.} {\bf 5}(2), 230 (2009). 


\bibitem{Han1}
P.G. Lind, M.C. Gonz\'{a}lez and H.J. Herrmann, {\it Phys. Rev. E}
{\bf 72}, 056127 (2005).

\bibitem{Han2}
P.G. Lind, H.J. Herrmann, {\it New Journal of Physics} {\bf 9}, 228
(2007).

\bibitem{Liu1}
J.-G. Liu, Y.-Z. Dang and Z.-T. Wang, {\it Physica A} {\bf 366}, 578
(2006).

\bibitem{Liu2}
J.-G. Liu, Z.-G. Xuan, Y.-Z. Dang, Q. Guo and Z.-T. Wang, {\it
Physica A} {\bf 377}, 302 (2007).

\bibitem{Zhou2009} T. Zhou, L. Lv and Y.-C. Zhang, arXiv:0901.0553v1.





\end{thebibliography}
\end{document}